\documentclass[12pt]{iopart}

\usepackage{iopams}
\usepackage{braket}
\usepackage{amsfonts}
\usepackage{color}
\usepackage[dvipsnames]{xcolor}
\usepackage{amssymb}
\usepackage{mathrsfs}
\usepackage{graphicx}
\usepackage{lmodern}
\usepackage{lineno}
\usepackage{hyperref}
\usepackage[normalem]{ulem}

\newcommand{\beq}{\begin{eqnarray}}
\newcommand{\eeq}{\end{eqnarray}}

\newcommand{\Real}{Re}
\newcommand{\Imag}{Im}

\def\keywords#1{\vspace{10pt}
     \begin{indented}
     \item[]\rm Keywords: #1\par
     \end{indented}}

\begin{document}



\title{Star product representation of coherent state path integrals}
\author{Jasel Berra--Montiel$^{1,2}$ }

\address{$^{1}$ Facultad de Ciencias, Universidad Aut\'onoma de San Luis 
Potos\'{\i} \\
Campus Pedregal, Av. Parque Chapultepec 1610, Col. Privadas del Pedregal, San
Luis Potos\'{\i}, SLP, 78217, Mexico}
\address{$^2$ Dual CP Institute of High Energy Physics, Mexico}

\eads{\mailto{\textcolor{blue}{jasel.berra@uaslp.mx}},\ 
}


\begin{abstract}
In this paper, we determine the star product representation of coherent path integrals. By employing the properties of generalized delta functions with complex arguments, the Glauber-Sudarshan P-function corresponding to a non-diagonal density operator is obtained. Then, we compute the Husimi-Kano Q-representation of the time evolution operator in terms of the normal star product. Finally, the optical equivalence theorem allows us to express the coherent state path integral as a star exponential of the Hamiltonian function for the normal product.  

\end{abstract}

\keywords{star porduct, quasi-probability distributions, path integral}


\section{Introduction}
The path integral quantization remains up to date one of the main tools for understanding quantum mechanics and quantum field theory; in particular, the formalism has proved to be extremly helpful to study perturbative approximations of a wide diversity of physical phenomena \cite{Kleinert}. The notion of path integration was extended to the complex plane with the introduction of coherent states \cite{Klauder1}, \cite{Baranger}, \cite{Kochetov}, motivating a prolific progress on the production of semiclassical methods focused on the analysis of non-integrable systems and quantum chaos \cite{Gutzwiller}. From another perspective, the phase space formulation of quantum mechanics, based on the early works of Wigner \cite{Wigner}, Weyl \cite{Moyal} and Moyal \cite{Moyal}  enabled to represent the quantum theory as a statistical theory defined on the classical phase space. Within this picture, the employment of coherent states have found a wide-ranging applications in quantum optics and quantum information \cite{Scully}, \cite{Schleich}. The main reason lies on the pioneering works proposed by Cahill and Glauber \cite{Cahill}, where a family of $s$-parametrized quasi-probability distributions was introduced in order to characterize non-classical effects by means of the overcomplete basis of coherent states, independently of the adopted ordering prescription. Later on, in 1978 P.~Sharan showed that the Feyman path integral, in the position and momentum representation, can be expressed as a Fourier transform of the star exponential for the Moyal product. This result was generalized to the case of scalar field theories in \cite{Dito}, by means of Berezin calculus and the c-equivalence of star products.

In this paper, we analyze the coherent state path integrals within the context of the star product quantization. By employing the properties of generalized delta functions with complex arguments, the Glauber-Sudarshan P-function corresponding to a non-diagonal density operator is obtained. Then, we compute the Husimi-Kano Q-representation of the time evolution operator in terms of the normal star product. Finally, the optical equivalence theorem allows us to express the coherent path integral as a star exponential for the normal product.  Determining this star product representation, by means of quasi-probability distributions, may allow us to analyze several aspects related to some inconsistences associated to regularization techniques and the continuum limit, encountered in coherent path integrals \cite{Wilson}. From another perspective, it will also be relevant to implement the formalism described here in order to establish the star product representation of spin foam integrals, in the context of loop quantum gravity and loop quantum cosmology, since recently, a family of quasi-probability distributions has been determined \cite{BM1}, \cite{BM2}, \cite{BM3}.

This paper is organized as follows, in section 2, we briefly review the basic construction of the coherent path integration. In section 3, by using the Glauber-Sudarshan and the Husimi-Kano quasi-probability distributions, the star product representation of the coherent state path integral is obtained. Finally, we introduce some concluding remarks in section 4.

\section{Path integral quantization in the coherent state representation}
In this section, we derive the matrix elements associated to the time evolution operator (in natural units with $\hbar=1$)
\begin{equation}
\hat{U}(T)=e^{-iT\hat{H}(\hat{a},\hat{a}^{\dagger})},
\end{equation}
by means of the path integral quantization technique within the coherent state representation, where $\hat{H}(\hat{a},\hat{a}^{\dagger})$ corresponds to the normal ordered Hamiltonian operator and $T=t_{f}-t_{i}$ stands for the total time lapse. Let us denote by $\ket{\alpha_{i}}$ and $\ket{\alpha_{f}}$ two arbitrary initial and final coherent states. Thus, the matrix element $\braket{\alpha_{f}|\hat{U}(T)|\alpha_{i}}$ can be expressed as
\begin{equation}\label{matrixel}
\braket{\alpha_{f}|e^{-iT\hat{H}(\hat{a},\hat{a}^{\dagger})}|\alpha_{i}}=\lim_{\epsilon\to 0\, N\to\infty}\braket{\alpha_{f}|\left( 1-i\epsilon \hat{H}(\hat{a},\hat{a}^{\dagger})\right)^{N}|\alpha_{i}},
\end{equation}
where, as usual, a partition of the time interval $T=t_{f}-t_{i}$, into $N$ infinitesimal time slices, each of lenght $\epsilon$, has been introduced. Now, following the formulation established, for example in \cite{Su}, let us insert an over-complete set of coherent states $\ket{\alpha}$, at each time $t_{j}$ of the partition, through the introduction of the completeness relation given, as an integral over the complex plane 
\begin{equation}
\int\frac{d^{2}\alpha}{\pi}\ket{\alpha}\bra{\alpha}=1,
\end{equation}
where $d^{2}\alpha=d\Real(\alpha)d\Imag(\alpha)$. The coherent states $\ket{\alpha}$ are defined as
\begin{equation}
\ket{\alpha}=\hat{D}(\alpha)\ket{0},
\end{equation} 
for $\alpha\in\mathbb{C}$, where $\hat{D}(\alpha)=\exp(\alpha\hat{a}^{\dagger}-\alpha^{*}\hat{a})$ corresponds to the displacement operator \cite{Scully}, and $\ket{0}$ stands for the vacuum ground eigenstate of a harmonic oscillator, such that $\hat{a}\ket{0}=0$. By expanding the exponential appearing in the diplacement operator $\hat{D}(\alpha)$, and using the Baker-Campbell-Hausdorff formula \cite{Hall} makes it clear that
\begin{equation}
\ket{\alpha}=\hat{D}(\alpha)\ket{0}=e^{-\frac{1}{2}|\alpha|^{2}}\sum_{n=0}^{\infty}\frac{\alpha^{n}}{\sqrt{n!}}\ket{n},
\end{equation} 
for $\ket{n}$ the number state satisfying $(\hat{a}^{\dagger})^{n}\ket{0}=\sqrt{n!}\ket{n}$. In addition,  the product of two coherent states proves to be
\begin{equation}
\braket{\beta|\alpha}=e^{-\frac{1}{2}|\alpha|^{2}-\frac{1}{2}|\beta|^{2}+\beta^{*}\alpha},
\end{equation}
which means that the coherent states are not orthogonal. With this information at hand, we compute that
\begin{eqnarray}
\mkern-50mu\braket{\alpha_{f}|\left( 1-i\epsilon \hat{H}(\hat{a},\hat{a}^{\dagger})\right)^{N}|\alpha_{i}}&=&\int\left( \prod_{j=1}^{N-1}\frac{d^{2}\alpha_{j}}{\pi}\right) \exp\left[-\frac{1}{2}|\alpha_{0}^{2}|-\frac{1}{2}|\alpha_{N}|^{2} \right. \nonumber \\ && \left. -\sum_{j=1}^{N-1}|\alpha_{j}|^{2}+\sum_{j=1}^{N}\alpha_{j}^{*}\alpha_{j-1}-i\sum_{j=1}^{N}\epsilon H(\alpha_{j-1},\alpha^{*}_{j})\right],  
\end{eqnarray}
where $H(\alpha_{j-1},\alpha^{*}_{j})=\braket{\alpha_{j}|\hat{H}(\hat{a},\hat{a}^{\dagger})|\alpha_{j-1}}$ is the function obtained from the normal ordered Hamiltonian operator $\hat{H}(\hat{a},\hat{a}^{\dagger})$ by working the substitutions $\hat{a}^{\dagger}\to\alpha^{*}_{j}$ and $\hat{a}\to\alpha_{j-1}$, and also we have designated the following boundary conditions $\alpha_{0}=\alpha_{i}$, $\alpha_{N}=\alpha_{f}$. 
By making use of the identity \cite{Novikov} 
\begin{equation}
\mkern-90mu\sum_{j=1}^{N}\alpha^{*}_{j}\alpha_{j-1}-\sum_{j=1}^{N-1}|\alpha_{j}|^{2}=\frac{1}{2}(\alpha^{*}_{N}\alpha_{N-1}+\alpha^{*}_{1}\alpha_{0})+\frac{\epsilon}{2}\sum_{j=1}^{N-1}\left(\alpha_{j}\frac{\alpha^{*}_{j+1}-\alpha^{*}_{j}}{\epsilon}-\alpha^{*}_{j}\frac{\alpha_{j}-\alpha_{j-1}}{\epsilon}\right), 
\end{equation}
the continuum limit $(\epsilon\to 0, N\to\infty)$ of the matrix element depicted in (\ref{matrixel}) reads (for further details see \cite{Su}, \cite{Novikov})
\begin{equation}\label{cspi}
\braket{\alpha_{f}|e^{-iT\hat{H}(\hat{a}^{\dagger},\hat{a})}|\alpha_{i}}=\int\mathcal{D}^{2}\left( \frac{\alpha}{\pi}\right)e^{\frac{1}{2}(|\alpha_{f}|^{2}-|\alpha_{i}|^{2})}e^{-\int_{t_{i}}^{t_{f}}dt\,\mathcal{L}(\alpha(t),\alpha^{*}(t))}, 
\end{equation}
where
\begin{equation}
\mathcal{L}(\alpha(t), \alpha^{*}(t))=\alpha^{*}(t)\dot{\alpha}(t)+iH(\alpha(t),\alpha^{*}(t)),
\end{equation}
and we have used the notation
\begin{equation}
\mathcal{D}^{2}\left( \frac{\alpha}{\pi}\right) =\lim_{N\to\infty}\prod_{j=1}^{N}\frac{d\Real(\alpha_{j})d\Imag(\alpha_{j})}{\pi},
\end{equation}
to denote the formal integration measure. As argued in \cite{Su}, these expressions comply consistency conditions according to the coherent state representation of transition amplitudes at zero temperature \cite{Itzykson}, \cite{Faddeev}, and  also fulfill the correct requirements stated by functional methods for calculating integrals in the complex plane. In case the functional integrals are Gaussians, the partition functions can be completely determined by means of the stationary phase method.  

\section{The Star product representation}
In order to construct the star product representation of the coherent state path integral, depicted in (\ref{cspi}), let us introduce the family of $s$-parametrized quasi-probability distributions in an integral form \cite{Scully}, \cite{Cahill}, given as
\begin{equation}\label{squasi}
F(\alpha,s)=\frac{1}{\pi^{2}}\int d^{2}\beta\, G(\beta,s)e^{\alpha\beta^{*}-\alpha^{*}\beta}
\end{equation}
where $G(\beta,s)$ denotes the $s$-ordered generalized characteristic function
\begin{equation}\label{G}
G(\beta,s)=\tr\left\lbrace \hat{D}(\beta)\hat{\rho}\right\rbrace e^{\frac{s}{2}|\beta|^{2}}.
\end{equation}
In the former expression, the operator $\hat{D}(\beta)$ corresponds to the displacement operator and $\hat{\rho}$ stands for the density operator of a quantum system. The parameter $s$ is related to the different ordering prescription of the operators $\hat{a}$ and $\hat{a}^{\dagger}$; for the value $s=1$ we obtain the Glauber-Sudarshan P-function (normal ordering) \cite{Sudarshan}, \cite{Klauder}, for $s=0$ we get the Wigner distribution (Weyl-symmetric ordering) \cite{Wigner}, and $s=-1$ determines the Husimi-Kano Q function (anit-normal ordering) \cite{Schleich}, \cite{Cahill} (for a generalzation to fields see \cite{BM}). 

Let us now consider the case of the non-diagonal density operator defined by $\hat{\rho}=\ket{\alpha_{i}}\bra{\alpha_{f}}$. For any operator $\hat{B}$, we can make use of the $s$-ordered quasi-probability distributions (\ref{squasi}) in order to express the matrix element $\braket{\alpha_{f}|\hat{B}|\alpha_{i}}$ as
\begin{eqnarray}\label{me}
\braket{\alpha_{f}|\hat{B}|\alpha_{i}}&=&\tr\left\lbrace \hat{B}\hat{\rho}\right\rbrace \nonumber\\
&=&\frac{1}{\pi^{2}}\int d^{2}\alpha\,d^{2}\beta\,tr\left\lbrace \hat{D}(\beta)\hat{B}\hat{\rho}\right\rbrace e^{\frac{s}{2}|\beta|^{2}+\alpha\beta^{*}-\alpha^{*}\beta}.
\end{eqnarray}
Since the path integral representation, analyzed in the previuous section, was determined through the normal ordering prescription, for the case $s=1$ the matrix element (\ref{me}) can be written as
\begin{eqnarray}\label{optical}
\braket{\alpha_{f}|\hat{B}|\alpha_{i}}&=&\int d^{2}\alpha\, P(\alpha)\braket{\alpha|\hat{B}|\alpha},\nonumber \\
&=&\int d^{2}\alpha\,P(\alpha)B_{Q}(\alpha,\alpha^{*}),
\end{eqnarray}
where $P(\alpha)$ corresponds to the Glauber-Sudarhan P-representation of the density operator $\hat{\rho}$, and $B_{Q}(\alpha,\alpha^{*})$ stands for the Husimi-Kano Q-representation of the operator $\hat{B}$ \cite{Scully}, \cite{Knight}. To be more precise, $P(\alpha)$ is a quasi-probability distribution of the form \cite{Mehta}:
\begin{equation}\label{Palpha}
P(\alpha)=\frac{e^{|\alpha|^{2}}}{\pi^{2}}\int d^{2}\beta\,\braket{-\beta|\hat{\rho}|\beta}e^{|\beta|^{2}+\beta^{*}\alpha-\beta\alpha^{*}},
\end{equation}
and the function $B_{Q}(\alpha,\alpha^{*})$ fulfills the properties associated to a probability distribution in  a sense of positibity, expressed as
\begin{equation}
B_{Q}(\alpha,\alpha^{*})=\braket{\alpha|\hat{B}|\alpha}.
\end{equation}
The equation (\ref{optical}) is known in the literature as the optical equivalence theorem \cite{Cahill}, \cite{Sudarshan}, and states that the expectation value of any normally ordered operator $\hat{B}$, is given by the average of the Q-representation of the operator $\hat{B}$ over the P-representation associated to the density operator $\hat{\rho}$.

Now, with the aim to obtain the star product representation of coherent path integrals, let us represent the matrix element $\braket{\alpha_{f}|e^{-iT\hat{H}(\hat{a},\hat{a}^{\dagger})}|\alpha_{i}}$  in terms of quasi-probabity distributions. By using the optical equivalence theorem (\ref{optical}), we see that
\begin{equation}\label{qme}
\braket{\alpha_{f}|e^{-iT\hat{H}(\hat{a},\hat{a}^{\dagger})}|\alpha_{i}}=\int d^{2}\alpha\, P(\alpha)\braket{\alpha|e^{-iT\hat{H}(\hat{a},\hat{a}^{\dagger})}|\alpha},
\end{equation}
where $\hat{H}(\hat{a},\hat{a}^{\dagger})$ corresponds to the normally ordered Hamiltonian operator, and $P(\alpha)$ denotes the Glauber-Sudarshan P-representation of the non-diagonal density operator $\hat{\rho}=\ket{\alpha_{i}}\bra{\alpha_{f}}$. In order to compute the integral (\ref{qme}), first it is necessary to analyze the Husimi-Kano Q-representation of the time evolution operator $\hat{U}(T)=e^{-iT\hat{H}(\hat{a},\hat{a}^{\dagger})}$. 
Given $\hat{B}(\hat{a},\hat{a}^{\dagger})$ and $\hat{C}(\hat{a},\hat{a}^{\dagger})$ two normally ordered operators, this means that we can express both as the normally ordered expansions
\begin{equation}
\hat{B}=\sum_{m,n=0}^{\infty}b_{mn}(\hat{a}^{\dagger})^{m}\hat{a}^{n}, \;\;\;\; \hat{C}=\sum_{p,q=0}^{\infty}c_{pq}(\hat{a}^{\dagger})^{p}\hat{a}^{q},
\end{equation}
with $b_{mn}, c_{pq}\in\mathbb{C}$. In terms of the expectation value over coherent states \cite{Lizzi} , we observe that
\begin{eqnarray}
\braket{\alpha|\hat{B}\hat{C}|\alpha}&=&\sum_{m,n,p,q}^{\infty}b_{mn}c_{pq}\braket{\alpha|(\hat{a}^{\dagger})^{m}\hat{a}^{n}(\hat{a}^{\dagger})^{p}\hat{a}^{q}|\alpha}, \nonumber \\
&=&\int \frac{d^{2}\beta}{\pi}\sum_{m,n,p,q}^{\infty}b_{mn}c_{pq}(\alpha^{*})^{m}\alpha^{q}\braket{\alpha|\hat{a}^{n}|\beta}\braket{\beta|(\hat{a}^{\dagger})^{p}|\beta}, \nonumber \\
&=& \int \frac{d^{2}\beta}{\pi}\sum_{m,n,p,q}^{\infty}b_{mn}c_{pq}(\alpha^{*})^{m}\alpha^{q}\beta^{n}(\beta^{*})^{p}\braket{\alpha|\beta}\braket{\beta|\alpha}, \nonumber \\
&=&\int \frac{d^{2}\beta}{\pi}B(\beta,\alpha^{*})C(\alpha,\beta^{*})|\!\braket{\alpha|\beta}\!|^{2} \label{istar}.
\end{eqnarray}
The expression (\ref{istar}) is known as the integral representation of the normal star product between the functions $B(\alpha,\alpha^{*}):=\braket{\alpha|\hat{B}|\alpha}$, $C(\alpha,\alpha^{*}):=\braket{\alpha|\hat{C}|\alpha}$, and is commonly denoted as $(B\star_{N}C)(\alpha,\alpha^{*})$. Alternatively, one can perform a series expansion, yielding the differential representation of the normal star product \cite{Hirshfeld}, \cite{Alexanian}
\begin{equation}
(B\star_{N} C)(\alpha,\alpha^{*})=B(\alpha,\alpha^{*})\exp\left\lbrace \frac{\overleftarrow{\partial}}{\partial \alpha}\frac{\overrightarrow{\partial}}{\partial \alpha^{*}}\right\rbrace C(\alpha,\alpha^{*}). 
\end{equation}  
which results completely equivalent to the expression (\ref{istar}). Considering that the evolution operator $\hat{U}(T)$, according to the spectral theorem \cite{Hall}, can be defined by a convergent power series
\begin{equation}
\hat{U}(T)=e^{-iT\hat{H}(\hat{a},\hat{a}^{\dagger})}=1-iT\hat{H}(\hat{a},\hat{a}^{\dagger})+\frac{(-iT)^{2}}{2!}\hat{H}^{2}(\hat{a},\hat{a}^{\dagger})+\cdots,
\end{equation}
then, the Husimi-Kano Q-representation of the evolution operator can be written as
\begin{equation}\label{Husimi}
\braket{\alpha|e^{-iT\hat{H}(\hat{a},\hat{a}^{\dagger})}|\alpha}=\exp_{\star_{N}}\left\lbrace -iTH(\alpha,\alpha^{*}) \right\rbrace,  
\end{equation}
where the normal star exponential is given by
\begin{equation}
\\exp_{\star_{N}}\left\lbrace -iTH(\alpha,\alpha^{*}) \right\rbrace=1-iTH(\alpha,\alpha^{*})+\frac{(-iT)^{2}}{2!}(H\star_{N}H)(\alpha,\alpha^{*})+\cdots.
\end{equation}

Our next task is to determine the Glauber-Sudarshan P-function associated to the non-diagonal density operator $\hat{\rho}=\ket{\alpha_{i}}\bra{\alpha_{f}}$. At first glance, this seems to be problematic since the P-function is usually related to the diagonal representation of the density operator in a coherent state basis \cite{Sudarshan}. These problems can be avoided by introducing the positive P-function \cite{Drummond}, which corresponds to a generalization of the P-representation and is obtained by doubling the dimension of the phase-space along with solving a Fokker-Planck type equation. For our purporses, however, it is more convenient to employ generalized delta functions to describe the non-diagonal parts of the denisty operator. 

A generalized delta function $\tilde{\delta}(z)$ is a delta function with complex arguments, symbolically represented as \cite{Poularkis},
\begin{equation}\label{gdf}
\tilde{\delta}(z)=\frac{1}{2\pi}\int_{-\infty}^{\infty}dx\,e^{-izx}
\end{equation}
such that
\begin{equation}
\int_{\infty}^{\infty}dx\,f(x)\tilde{\delta}(x-z)=f(z),
\end{equation}
where $z\in\mathbb{C}$. The test function $f(z)$ is analytic on the complex variable $z$, and also must decay sufficiently rapidly at large distances along the real axis, that is, $f(x+iy)$ is of order $O(|x|^{-N})$ as $|x|\to\infty$ for all $N$. As described in \cite{Brewster} the idea of introducing generalized delta functions is to extend the ordinary Dirac delta function to the complex plane by making use of contour integrals (see \cite{Poularkis}, \cite{Brewster}, for more details). 
One of the most compelling properties of the generalized delta function involves separate integrations over real and imaginary parts of a given function, that is
\begin{eqnarray}
\int_{-\infty}^{\infty}d\Real(\alpha)\,f(\Real(\alpha),\Imag(\alpha))\tilde{\delta}(\Real(\alpha)-z)&=&f(z,\Imag(\alpha)),\label{gd1}\\
\int_{-\infty}^{\infty}d\Imag(\alpha)\,f(\Real(\alpha),\Imag(\alpha))\tilde{\delta}(\Real(\alpha)-z)&=&f(z,\Real(\alpha))\label{gd2},
\end{eqnarray}
here $\alpha\in\mathbb{C}$ and $\Real(\alpha), \Imag(\alpha)$ stands for its real and imaginary parts respectively. These formulas has previuously been applied in the positive P-representation to determine the Glauber-Sudarshan P-function associated to a cat state \cite{Drummond}, \cite{Brewster}.

With the preceding formulas, we now compute the P-representation corresponding to the non-diagonal density operator $\ket{\alpha_{i}}\bra{\alpha_{f}}$. By using the explicit expression for $P(\alpha)$ in (\ref{Palpha}), and the definition of the generalized delta function (\ref{gdf}), we obtain
\begin{eqnarray}\label{Pfunction}
P(\alpha)&=&\frac{e^{|\alpha|^{2}}}{\pi^{2}}\int d^{2}\beta\,\braket{-\beta|\alpha_{i}}\braket{\alpha_{f}|\beta}e^{|\beta|^{2}+\beta^{*}\alpha-\beta\alpha^{*}}, \nonumber \\
&=&\tilde{\delta}\left(\frac{\alpha_{i}+\alpha_{f}^{*}}{2}-\Real(\alpha)\right)\tilde{\delta}\left(i\frac{\alpha_{i}-\alpha_{f}^{*}}{2}-\Imag(\alpha)\right). 
\end{eqnarray}
Finally, we are ready to obtain the star product representation of the matrix element $\braket{\alpha_{f}|e^{-iT\hat{H}(\hat{a}^{\dagger},\hat{a})}|\alpha_{i}}$. By substituting the Husimi-Kano representation of the evolution operator (\ref{Husimi}) and the Glauber-Sudarshan P-function of the density operator (\ref{Pfunction}) into the expression (\ref{qme}), using the properties of the generalized delta function (\ref{gd1}), (\ref{gd2}), we have
\begin{equation}\label{starpath}
\braket{\alpha_{f}|e^{-iT\hat{H}(\hat{a},\hat{a}^{\dagger})}|\alpha_{i}}=\braket{\alpha_{f}|\alpha_{i}}\left.\exp_{\star_{N}}\left\lbrace -iTH(\alpha,\alpha^{*}) \right\rbrace\right|_{\alpha=\alpha_{i}, \alpha^{*}=\alpha_{f}^{*}}.  
\end{equation}
Since the coherent path integral is given by the matrix element (\ref{cspi}), its corresponding star product representation follows consequently from equation (\ref{starpath}). This means, that the path integral for coherent states is realized by the normal star exponential of the classical Hamiltonian function. It is worthwhile to mention that if we instead choose $\braket{\alpha_{f}|\alpha_{i}}=e^{\alpha_{f}^{*}\alpha_{i}}$, as the inner product between coherent states, the formula (\ref{starpath}) agrees with the expression found in \cite{Dito}, where the star-quantization of the free scalar field was introduced by means of Berezin calculus and the c-equivalence between the Moyal and normal star products in the holomorphic representation. However, since the formalism presented here is developed by means of the analysis of $s$-parametrized quasi-probability distributions, it may demonstrate relevant in order to explore the repercussion of changing the ordering on the path integral by continuously varying the parameter $s$ without making use of the c-equivalence of star products, which in some cases may result quite demanding.

\section{Conclusions}
\label{sec:conclu}
 
In this paper, the star product representation of coherent state path integrals was determined. In particular, by employing the properties of generalized delta functions with complex arguments, we obtain the Glauber-Sudarshan P-function associated to a non-diagonal density operator. Then, we compute the Husimi-Kano Q-representation of the time evolution operator in terms of the normal star product. Finally, the optical equivalence theorem allows us to express the coherent path integral in terms of the star exponential of the classical Hamiltonian for the normal product. We claim that our developments will be helpful to analyze several inconsistencies related to regularization and continuum limits, encountered in coherent state path integration \cite{Wilson}. From a different perspective, it will also be relevant to implement the techniques described here to establish the star product representation of spin foam models in the loop quantum gravity and loop quantum cosmology framework, since recently, a family $s$-parametrized quasi-probability distributions and its relation with the Wigner-Weyl representation has been obtained \cite{BM1},\cite{BM2},\cite{BM3}. We intend to dedicate a future publication to address this work in progress.

\section*{Acknowledgments}
The author would like to acknowledge financial support from CONACYT-Mexico
under the project CB-2017-283838.

\section*{References}

\bibliographystyle{unsrt}

\end{document}